\documentclass[fleqn,twoside]{article}

\usepackage{graphicx}
\usepackage[figuresright]{rotating}
\usepackage{amsmath}
\usepackage{amssymb}
\usepackage{espcrc2}
\usepackage{mcite}
\newcommand{\sign}[0]{\text{sign}}
\newcommand{\bra}[1]{\left\langle #1\right|}
\newcommand{\ket}[1]{\left|#1\right\rangle}
\newcommand{\psibeta}[0]{\psi^{l}}
\newcommand{\ketbra}[0]{\ket{\psibeta}\bra{\psibeta}}
\newcommand{\lambdabeta}[0]{\lambda_{l}}

\newcommand{\Tr}[0]{\text{Tr}}

\title{Dynamical overlap simulations using HMC}

\author{N. Cundy\address[PW]{Department of Physics, Universit\"at
    Wuppertal, Gaussstrasse 19, Wuppertal, Germany.}\thanks{Speaker}
\thanks{NC is supported by EU 
        grant MC-EIF-CT-2003-501467}, %
        S. Krieg\addressmark[PW],  %
        A. Frommer\address[MW]{Department of Mathematics, Universit\"at
    Wuppertal.},
	Th. Lippert\address[J]{John von Neumann Institute for Computing,
	  J\"ulich Research
	  Centre, 52425 J\"ulich, Germany.},
	and K. Schilling\addressmark[PW].
}
       
\begin{document}

\begin{abstract}
We apply the Hybrid Monte Carlo method to the simulation of overlap fermions.
We give the
fermionic force for the molecular dynamics update. We present early results on a small
dynamical chiral ensemble.
\end{abstract}

\maketitle

\section{INTRODUCTION}
It is impossible to run dynamical simulations at realistic masses with
Wilson fermions, which explicitly break chiral
symmetry. It is still uncertain whether taking the fourth root of the
determinant in staggered fermion simulations violates locality. However,
overlap fermions satisfy a lattice chiral symmetry, and have no such
conceptial difficulties. The disadvantage with overlap fermions is
that the calculation of the matrix
sign function used in the overlap operator requires O(100) calls to
the Wilson operator: overlap fermions are computationally costly! We
have developed methods which can accelerate the inversion of the
overlap operator by more than a factor of 4~\cite{paperII, paperIII, stefan}, bringing
simulations on small lattices and at large masses to within the
capabilities of modern computers. In
this article, we sketch the development of a Hybrid Monte Carlo~\cite{kennedy} algorithm for dynamical overlap
fermions~\cite{paper4} (see also ~\cite{FODOR,kalman}).
\section{FERMIONIC FORCE}\label{sec2}
The massive overlap operator is given by
\begin{gather}
D = (1+\mu) + (1-\mu)\gamma_5 \text{sign}(\gamma_5 D_W),\nonumber
\end{gather}
where $D_W$ is the Wilson operator with a negative mass parameter, and the
bare fermion mass is proportional to
$\mu/(1-\mu)$. 
The fermionic force is the differential of $1/(D^{\dagger}D)$ with
respect to the gauge field $U_{\mu}(x)$.
The calculation of the matrix sign function has to be accelerated by projecting
out the smallest Wilson eigenvectors $\ket{\psi^l}$ (with eigenvalues $\lambda^l$) and treating them
exactly. The eigenvectors can be differentiated using a procedure
analogous to first order quantum mechanics perturbation theory. The
fermionic force acting on a momentum $\Pi$ conjugate to a spinor field
$\phi$ is $F_{\mu}(x):$
\begin{align}
&{\frac{d\sign(Q)}{dU_{\mu}(x)}}_{nm} = \kappa \frac{1}{a}
\omega_{\eta}A^{k}_{nb}\bigg[\frac{1}{a^2}Q_{bx}\gamma_5(1-\gamma_{\mu})\delta_{x+\mu,c}Q_{cd}\nonumber\\
&-\zeta_{k}\gamma_5(1-\gamma_{\mu})\delta_{b,x}\delta_{x+\mu,d}\bigg]A^{k}_{de}
\left(1-\ketbra\right)_{em}\nonumber\\
&-\kappa  \left(\frac{\frac{1}{a}Q\omega_{k}}{\frac{1}{a^2}Q^2
  +\zeta_{k}}\right)_{nb}P_{l
  bx}\gamma_5
\nonumber\\&\phantom{lots of nice space}
(1-\gamma_{\mu})\delta_{x-e_{\mu},c}\left(\ketbra\right)_{cm}\nonumber\\
&-\kappa  \left(\frac{\frac{1}{a}Q\omega_{k}}{\frac{1}{a^2}Q^2
  +\zeta_{k}}\right)_{nb}\left(\ketbra\right)_{bx}\gamma_5
\nonumber\\&\phantom{a great deal of much space}
(1-\gamma_{\mu})\delta_{x,c+e_{\mu}}P_{l cm}\nonumber\\
&+\kappa P_{l
  nx}\gamma_5\epsilon(\lambdabeta)(1-\gamma_{\mu})\delta_{x,c+e_{\mu}}
\left(\ketbra\right)_{cm}\nonumber\\
&+\kappa
\left(\ketbra\right)_{nx}\gamma_5(1-\gamma_{\mu})\delta_{x,c+e_{\mu}}
\epsilon(\lambdabeta)P_{l
  cm}\nonumber\\
&-\left(\ketbra\right)_{nm}
{\frac{d\epsilon(\lambdabeta)}{d\lambdabeta}}
\nonumber\\&\phantom{some more space}
\bra{\psibeta}_x\gamma_5(1-{\gamma_{\mu}})\delta_{x,c+e_{\mu}}\ket{\psibeta}_c\nonumber\\
&F_{\mu}(x) =
-\left(\phi^{\dagger}\frac{1}{D^{\dagger}D}\right)_n\left\{\gamma_5,{\frac{d\sign(Q)}{dU_{\mu}(x)}}\right\}_{nm}
\nonumber\\
&\phantom{a lot of nice big space fgh}(1-\mu^2)\left(\frac{1}{D^{\dagger}D}\phi\right)_m,
\end{align}
where $\epsilon$ is the sign function and 
\begin{gather}
P_{l} = (Q-\lambdabeta)^{-1}(1-\ketbra).
\end{gather}
 The delta function in the fermionic
force has to be treated exactly, because ignoring it will lead to
large jumps in the hybrid Monte Carlo energy, and an unacceptably
small acceptance rate. This can be done by adding the
following terms to the leapfrog update:
\begin{align}
&U_C =
e^{i\tau_c\Pi_-(\Delta\tau/2)}U_-(\Delta\tau/2),\nonumber\\
&\Pi^N_+(\Delta\tau/2) =
\Pi^N_-(\Delta\tau/2)\sqrt{1+\frac{4d}{{(\Tr(\Pi_-^N))}^2}},\label{momupdate}\\
&U_+(\Delta\tau/2) =
e^{-i\tau_c\Pi_+(\Delta\tau/2)}U_C,\\
&d =  \text{Re}(\bra{\phi}\frac{1}{D^{\dagger}D(U_{C-})}
\gamma_5\ket{\psi}
\bra{\psi}\frac{1}{D^{\dagger}D(U_{C-})} \ket{\phi}),
\end{align}
where $U$ is the gauge field, $\tau$ the molecular dynamics artificial time, $U_C$ the gauge field at the
moment of the crossing (which can be calculated to numerical precision
using a Newton-Raphson procedure), $\Pi^N$ the component of the
crossing normal the $\lambda=0$ surface, and the $+$ sign
indicates the gauge field/momentum after the crossing, i.e. after the
smallest eigenvalue has changed sign. This procedure is area
conserving, reversible, and conserves energy up to $O(\Delta\tau)$. A simple
extension of this method will conserve energy up to
$O(\Delta\tau^2)$~\cite{paper4}. 

We cannot use the momentum update (\ref{momupdate}) when
$1+{4d}/{{(\Tr(\Pi_-^N))}^2}<0$. The authors of ~\cite{FODOR}
suggested that in this case, we reflect the momentum of the potential
wall, setting $\Pi_+^N = -\Pi^N_-$ (a useful analogy is a
classical mechanics particle approaching a potential wall). We must use this
``reflection algorithm'' for all $d<-{(\Tr(\Pi_-^N))}^2/4$ (and there will be no change in
the topological charge), and the ``transmission algorithm'' outlined
above for $d>-{(\Tr(\Pi_-^N))}^2/4$ (which will lead to a change in the topological charge).  

\section{CHIRAL PROJECTION}
It was suggested in a study of the Schwinger model~\cite{BODE} that
some gain could be achieved by projecting into the chiral sector with
no zero modes (the zero modes have to be accounted for by re-weighting
when the ensemble averages are taken). Because
$[D^{\dagger}D,\gamma_5] = 0$, we can use the operators
$D^{\dagger}D P_{\pm}$ (with $P_{\pm} = \frac{1}{2}(1\pm \gamma_5)$) to reconstruct the entire non-zero
eigenvalue spectrum of the overlap operator. One can show that
$\det D = \mu^{|Q|}\det (D^{\dagger}DP_{\pm})$, where $Q$ is
the topological charge, and the sign
is chosen so that we work in the chiral sector without zero modes. Because
$D^{\dagger}DP_{\pm}$ requires only one call to the sign
function, we can use this to generate a single flavour ensemble in half
the time we could generate a two flavour ensemble. This method has two advantages: it will allow more
frequent topological charge changes, reducing the autocorrelation length, and secondly there
is no exceptionally large fermionic force when there is a zero
mode. However, it will also generate a large number of $Q\neq 0$
configurations which will be weighted by a small factor when we take
the ensemble average. To preserve detailed balance, we cannot allow
the gauge field to move into the ``wrong'' chiral sector during the
molecular dynamics~\cite{paper4}. 
\begin{figure}
\begin{tabular}{c }
 \includegraphics[width = \columnwidth,height =
  4.9cm]{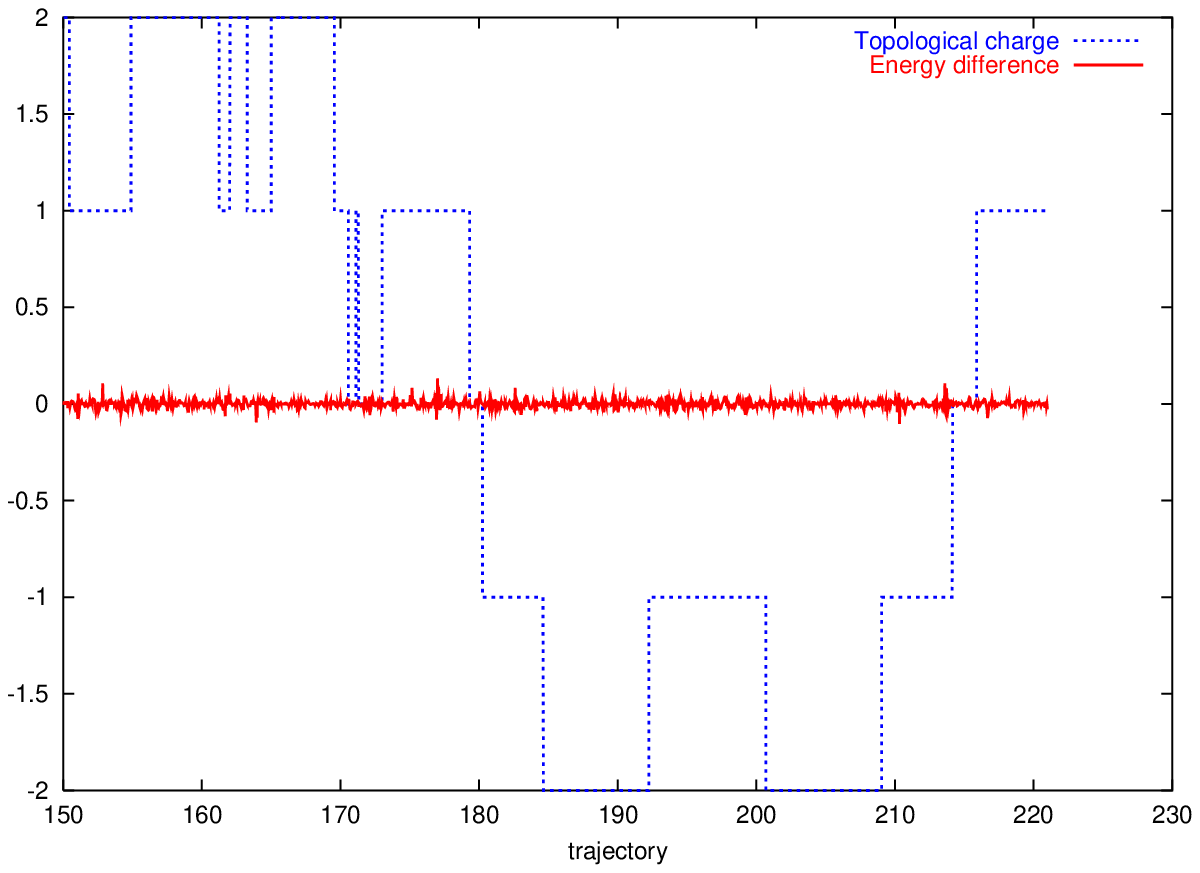}\\
\includegraphics[width = \columnwidth,height =
  4.9cm]{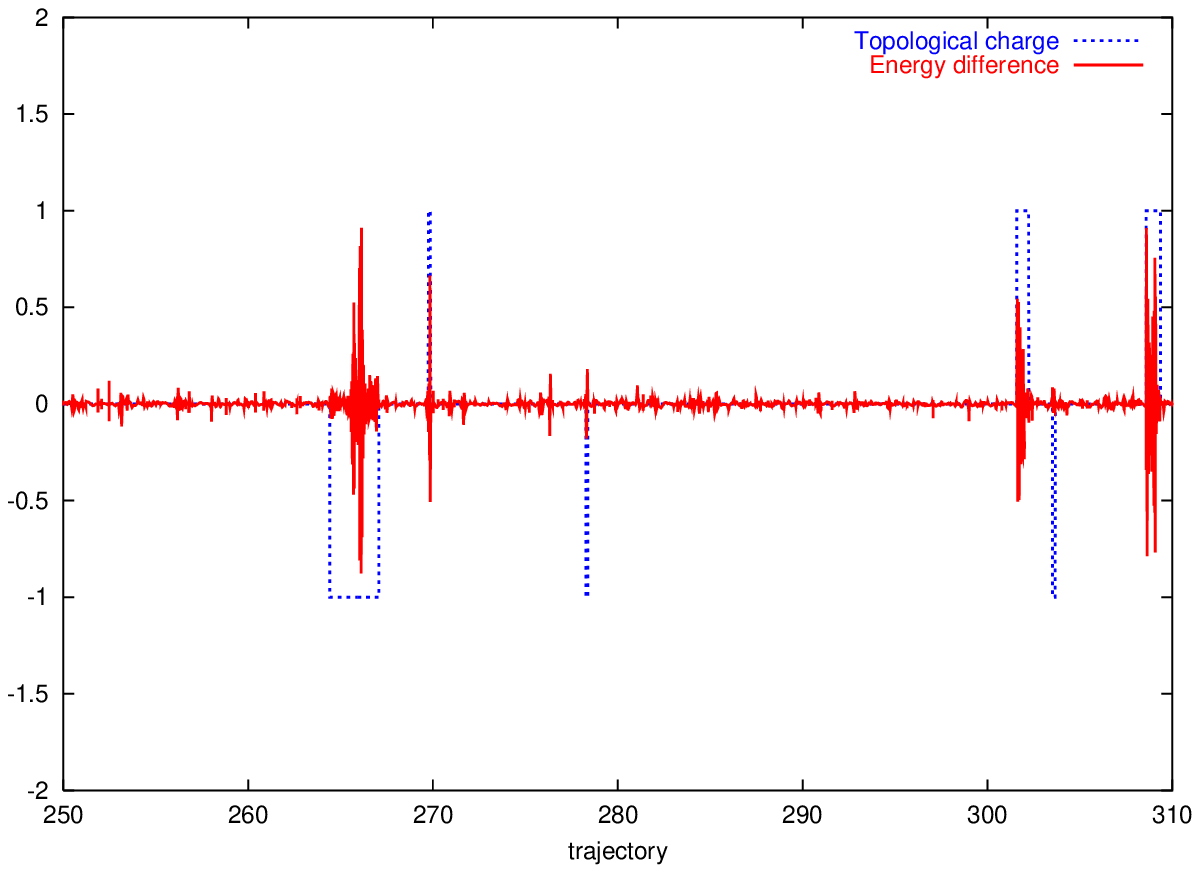}
\end{tabular}
\caption{Energy conservation plots for the $4^4$ $\beta = 5.4$ $\kappa
  = 0.225$, $\mu = 0.05$ ensembles, with (top) and without (bottom)
  chiral projection.}\label{fig1}
\end{figure}
\section{NUMERICAL TESTS}
We tested the algorithm described in section \ref{sec2} on $4^4$
$\beta = 5.4$ lattices with masses $\mu = 0.05,0.1,0.2,0.3,0.4$ and
$0.5$. Figure \ref{fig1} plots the topological charge and $dE/d\tau$
($E$ is the HMC energy) for each molecular dynamics time step for
 70 trajectories taken from the $\mu = 0.05$ ensembles with and
without chiral projection. There are no large
spikes in the energy difference, showing that our correction step does
indeed conserve energy. Secondly, the chiral projected ensemble, as
expected, has more frequent topological charge changes, and a
considerably smaller fermionic force when $Q\neq 0$.
\begin{table}
\begin{tabular}{l l|l l l}
$\mu$&$m_b$&$<Q^2>$ &$P_l$&$<S_{pf}>$\\
\hline
0.05&0.19
&0.016(1)&0.061(21)&0.578(2)
\\
0.1&0.40
&0.09(3)&0.064(18)&0.575(2)
\\
0.2&0.89
&0.16(3)&0.048(10)&0.568(2)
\\
0.3&1.52
&0.24(4)&0.027(9)&0.562(1)
\\
0.4&2.37
&0.29(4)&0.032(9)&0.561(1)
\\
0.5&3.56
&0.32(10)&0.005(10)&0.554(1)
\\
\hline
0.05&0.19
&0.01(1)&0.063(6)&0.576(1)
\\
0.1&0.40
&0.13(3)&0.071(7)&0.574(1)
\\
0.2&0.89
&0.29(4)&0.043(6)&0.569(1)
\\
0.3&1.52
&0.34(5)&0.048(8)&0.562(1)
\\
0.4&2.37
&.40(5)&0.029(9)&0.559(1)
\\
0.5&3.56
&0.45(14)&0.032(13)&0.554(1)
\end{tabular}
\caption{The plaquette ($S_pg$), Polykov loop ($P_l$) and ensemble average of $Q^2$ for the $4^4$
  $\kappa = 0.225$ ensembles with (top) and without (bottom) chiral projection.}\label{tab:tab1}
\end{table}

Table \ref{tab:tab1} gives the plaquette, Polykov loop, and topological
susceptibility for our ensembles. The plaquettes and Polykov loops are in good agreement
between the chiral projected and non-chiral projected ensembles; the
average values of $Q^2$ (proportional to the topological susceptibility) are not, for reasons which we will
outline in a later paper~\cite{paper4}. The general behaviour of the
plaquette and topological susceptibility are in agreement with our
expectations at these large masses (our smallest mass is of the same
order of magnitude as the strange quark mass), and the Polykov loop is
small, suggesting that our configurations are confined. 
\section{CONCLUSIONS}
We have shown that it is possible to run dynamical simulations using
overlap fermions, and that the delta function in the fermionic force
can be overcome. The results on our small test configurations are
sensible. We hope to begin simulations on $12^4$ and similar lattices
in the next year.
\bibliographystyle{elsart-num} 
\bibliography{proceedings_me}

\end{document}